\documentclass[11pt,letterpaper]{article}

\usepackage[margin=1in]{geometry}
\usepackage[T1]{fontenc}
\usepackage{microtype}
\usepackage{parskip}
\usepackage{setspace}
\usepackage[sc]{mathpazo}        
\usepackage[scale=0.95]{tgheros} 
\usepackage{sectsty}

\usepackage{graphicx}
\usepackage[table]{xcolor}
\usepackage{booktabs}
\usepackage{tabularx}
\usepackage{longtable}
\usepackage{array}
\usepackage[export]{adjustbox}   
\usepackage{makecell}
\usepackage{float}

\usepackage{caption}
\usepackage[backend=biber,style=numeric,sorting=none,doi=true,url=true]{biblatex}
\usepackage{url}             
\usepackage[colorlinks=true]{hyperref}
\hypersetup{
  colorlinks=true,
  linkcolor=blue!50!black,
  urlcolor=blue!50!black,
  citecolor=blue!50!black,
  pdfauthor={Shams El-Adawy et al.},
  pdftitle={Experimental Skills for Non-PhD Roles in the Quantum Industry}
}

\usepackage{titletoc}
\titlecontents{section}
  [0em]{\vspace{2pt}}{\contentslabel{1.5em}}{}{\titlerule*[.5pc]{.}\contentspage}
\titlecontents{subsection}
  [1.5em]{\vspace{1pt}}{\contentslabel{1.8em}}{}{\titlerule*[.5pc]{.}\contentspage}

\usepackage{enumitem}
\setlist{nosep,leftmargin=1.2em}

\usepackage[most]{tcolorbox}
\newtcolorbox{takeawaybox}{
  colback=blue!5!white, colframe=blue!40!black,
  boxrule=0.8pt, arc=2mm, left=6pt, right=6pt, top=6pt, bottom=6pt,
  title=\sffamily\bfseries Main Takeaways
}
\newtcolorbox{questionbox}{
  enhanced, breakable,
  colback=gray!8, colframe=gray!50,
  borderline west={3pt}{0pt}{gray!60},
  boxrule=0.5pt, arc=2mm,
  left=8pt, right=8pt, top=8pt, bottom=8pt,
  fonttitle=\sffamily\bfseries, coltitle=black,
  title={\sffamily\bfseries Key Question}
}

\AtBeginEnvironment{quote}{\small\itshape}

\usepackage{pagecolor}

\definecolor{brandnavy}{HTML}{243B7A}
\sectionfont{\sffamily\bfseries\color{brandnavy}}
\subsectionfont{\sffamily\bfseries\large\color{brandnavy}}
\subsubsectionfont{\sffamily\bfseries\normalsize\color{brandnavy}}
\usepackage{titlesec}
\titleformat{\paragraph}[runin]{\sffamily\bfseries\color{brandnavy}}{\theparagraph}{0.5em}{}[]
\titleformat{\subparagraph}[runin]{\sffamily\bfseries\color{brandnavy}}{\thesubparagraph}{0.5em}{}[]

\arrayrulecolor{brandnavy!35} 

\definecolor{tabheadbg}{RGB}{236, 240, 247} 
\definecolor{tabstripe}{RGB}{248, 250, 253} 
\definecolor{tabskillbg}{RGB}{243, 246, 252} 

\newcolumntype{C}[1]{>{\centering\arraybackslash}p{#1}}
\newcolumntype{S}[1]{>{\columncolor{tabskillbg}\centering\arraybackslash}p{#1}}

\usepackage{fancyhdr}
\renewcommand{\headrulewidth}{0.2pt}

\makeatletter
\renewcommand{\headrule}{%
  \vspace{-3pt}\hbox to\headwidth{\color{brandnavy}\leaders\hrule height \headrulewidth\hfill}}
\makeatother

\newcolumntype{P}[1]{>{\raggedright\arraybackslash}p{#1}}

\providecommand{\seriesfontsize}{\normalsize}
\providecommand{\seriesitem}[3]{}
\providecommand{\seriesitemcurrent}[3]{}
\providecommand{\seriesstripe}[5]{}

\renewcommand{\seriesfontsize}{\large} 
\newcommand{\serieslinesep}{12pt}       

\renewcommand{\seriesitem}[3]{%
  \noindent
  \begin{minipage}[t]{0.72\linewidth}
    {\sffamily\seriesfontsize\color{white} Report #1:\enspace #2}%
  \end{minipage}%
  \begin{minipage}[t]{0.28\linewidth}\raggedleft
    {\sffamily\seriesfontsize\color{white!85} #3}%
  \end{minipage}\par\vspace{\serieslinesep}%
}

\renewcommand{\seriesitemcurrent}[3]{\seriesitem{#1}{#2}{#3}}

\renewcommand{\seriesstripe}[5]{%
  \begin{tcolorbox}[enhanced, breakable,
    colback=brandnavy, colframe=brandnavy,
    boxrule=0pt, sharp corners,
    left=0pt, right=0pt, top=4pt, bottom=10pt,
    borderline south = {1pt}{0pt}{white!80}
  ]
    \noindent
    \begin{minipage}[t]{0.62\linewidth}
      {\sffamily\small\color{white!85}\MakeUppercase{#1}}%
    \end{minipage}%
    \begin{minipage}[t]{0.38\linewidth}\raggedleft
      {\sffamily\small\color{white!85} Publication date}%
    \end{minipage}

    \par\vspace{3pt}{\color{white!70}\rule{\linewidth}{0.4pt}}\vspace{4pt}

    \seriesitemcurrent{#2}{#4}{#3}

    {\sffamily\seriesfontsize\color{white!85} Upcoming in the series:}\par\vspace{2pt}
    #5
  \end{tcolorbox}%
}

\newcommand{\upcomingseriesitems}{%
  \seriesitem{2}{Categorization of Roles in the Quantum Industry}{TBD}%
  \seriesitem{3}{Profiles of Roles in the Quantum Industry}{TBD}%
}
\newcommand{\tocsection}[1]{%
  \phantomsection
  \addcontentsline{toc}{section}{#1}%
  \section*{#1}%
}

\newcommand{\logorowscale}{0.90}
\newcommand{\logospace}{1.2cm}

\title{%
  \vspace{-1.2cm}%
  \resizebox{\logorowscale\textwidth}{!}{%
    \makebox[\textwidth]{%
      \adjincludegraphics[valign=c,height=0.66cm]{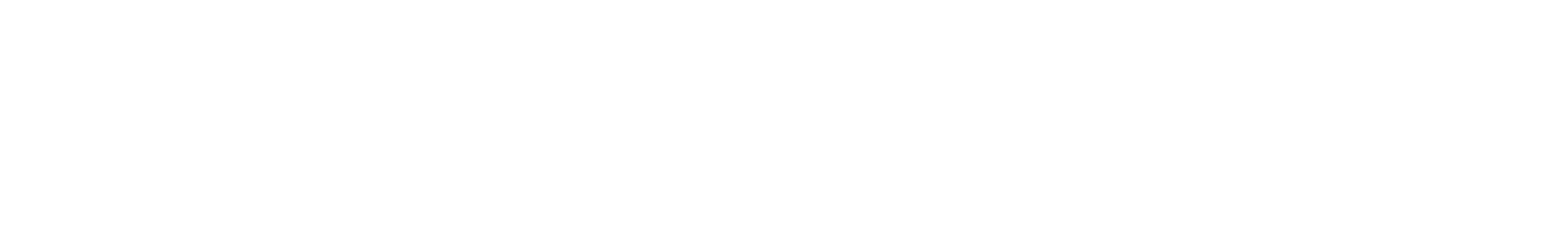}%
      \hspace{\logospace}%
      \adjincludegraphics[valign=c,height=0.60cm]{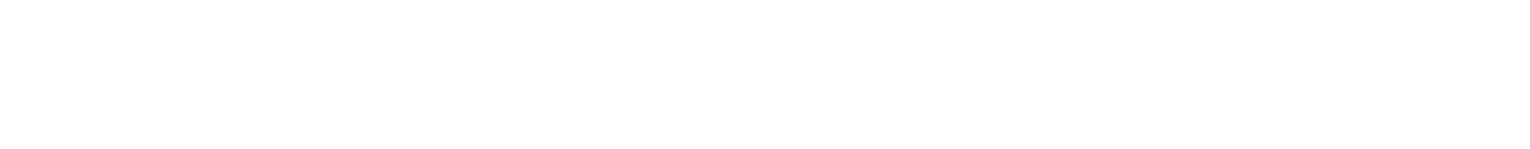}%
    }%
  }\\[1.2cm]
  {\sffamily\bfseries\Huge Experimental Skills for Non-PhD Roles in the Quantum Industry}\\[0.6cm]
  {\sffamily\large Report 1 — October 2025}
}

\author{%
  \parbox{\textwidth}{\centering
    \begingroup
    \setlength{\tabcolsep}{18pt}\renewcommand{\arraystretch}{1.2}%
    \begin{tabular*}{0.9\textwidth}{@{\extracolsep{\fill}} c c}
      \makecell{\textbf{Shams El-Adawy}\\ \textbf{Heather J. Lewandowski}\\[2pt]\small University of Colorado Boulder} &
      \makecell{\textbf{A.\,R. Pi\~{n}a}\\ \textbf{Benjamin M.\,Zwickl}\\[2pt]\small Rochester Institute of Technology}
    \end{tabular*}
    \endgroup
  }%
}

\date{}

\begin{document}

\begingroup
  \pagecolor{brandnavy}\color{white}
  \hypersetup{linkcolor=cyan!25, urlcolor=cyan!25, citecolor=cyan!25}

  \maketitle

  \seriesstripe{Quantum Workforce Report Series}{1}{October 2025}{Experimental Skills for Non-PhD Roles in the Quantum Industry}{\upcomingseriesitems}

  \thispagestyle{empty}
\endgroup

\clearpage
\nopagecolor
\color{black}
\hypersetup{linkcolor=blue!50!black, urlcolor=blue!50!black, citecolor=blue!50!black}

\tableofcontents
\thispagestyle{empty}
\clearpage


\tocsection{Executive summary}

As the quantum information science and engineering (QISE) workforce grows, there is an anticipated need for professionals with bachelor’s and master’s degrees who can fill a wide range of roles in the quantum industry \cite{el2025industry}. This report identifies the \textbf{experimental skills} needed for individuals with bachelor’s or master’s degrees to succeed in quantum industry roles. Through semi-structured interviews with quantum industry employers, we gathered data on \textbf{22 distinct positions} spanning hardware, software, and business functions. While employers describe varying expectations of quantum expertise, the unifying requirement across these roles is proficiency in experimental skills, which fall into four key categories: 
\begin{enumerate}
  \item \textbf{Instrumentation}
  \item \textbf{Computation and data analysis}  
  \item \textbf{Experimental and project design}  
  \item \textbf{Communication and collaboration} 
\end{enumerate}

Positions open to bachelor’s and master’s graduates use all four skill areas, but the balance of experimental skill set needed differs. Bachelor’s roles lean toward \textbf{instrumentation}, \textbf{computation and data analysis}, as well as \textbf{experimental and project design} skills. Individuals in these roles build, operate, and troubleshoot hardware, and they gather and interpret data to design and carry out experiments. Master’s roles stand out with the \textbf{communication and collaboration} skills needed on top of the other three skill categories. Individuals in these roles oversee experiments, coordinate teams, and align efforts with company and client needs. By articulating experimental skills needed for bachelor’s and master’s roles in the quantum industry, this report provides actionable insights for educators developing QISE courses and programs.

\vspace{0.5cm} \textbf{Recommendations:} \begin{enumerate}

   \item \textbf{Prioritize the discussion of hardware in QISE theory courses}: Undergraduate programs should emphasize the theoretical foundations of quantum hardware within QISE courses  to better prepare students for entry-level positions in the quantum industry. 
   Currently, many introductory QISE courses focus solely on abstract theoretical concepts with limited time spent on theory underlying the experiments and devices \cite{pina2025landscape}. 
   Thus, shifting the focus of these courses to include discussions on theory behind hardware components will better align QISE education with industry needs. 
   \item \textbf{Increase experimental training through instructional labs}: Many of the experimental skills needed for non-PhD positions in the quantum industry overlap with general experimental skills that are valuable across a range of career paths. Many of these  skills are already part of the learning goals of existing lab courses, which makes this recommendation achievable with minimal additional resources. Increasing opportunities to engage in experimental skill development within existing labs would better prepare students for positions in the quantum industry, while simultaneously strengthening preparation for other career paths.
   \item \textbf{Include professional skills in undergraduate QISE education}: Employers consistently identify communication and collaboration as critical skills. These skills should be embedded within QISE courses and programs to ensure students are prepared for industry roles. \end{enumerate}

\tocsection{Motivation}

The quantum industry is growing with increasing need for talent at all educational levels. While PhD-level expertise remains important for research and development, there is an anticipated need for quantum industry professionals with bachelor’s and master’s degrees \cite{el2025industry}. These roles are projected to be critical for the deployment of quantum technologies at scale. Despite this projected need, the specific skills required for these non-PhD roles remain underdefined, which makes it challenging to design courses and programs aligned with workforce needs \cite{el2025insights}. Understanding which experimental skills employers desire can help align curricula with workforce needs.

\tocsection{Data collection and analysis}

We conducted 37 semi-structured interviews with quantum industry employees based in the United States between December 2024 and May 2025. 

\begin{figure}[h!]
  \centering
  \includegraphics[width=0.9\linewidth]{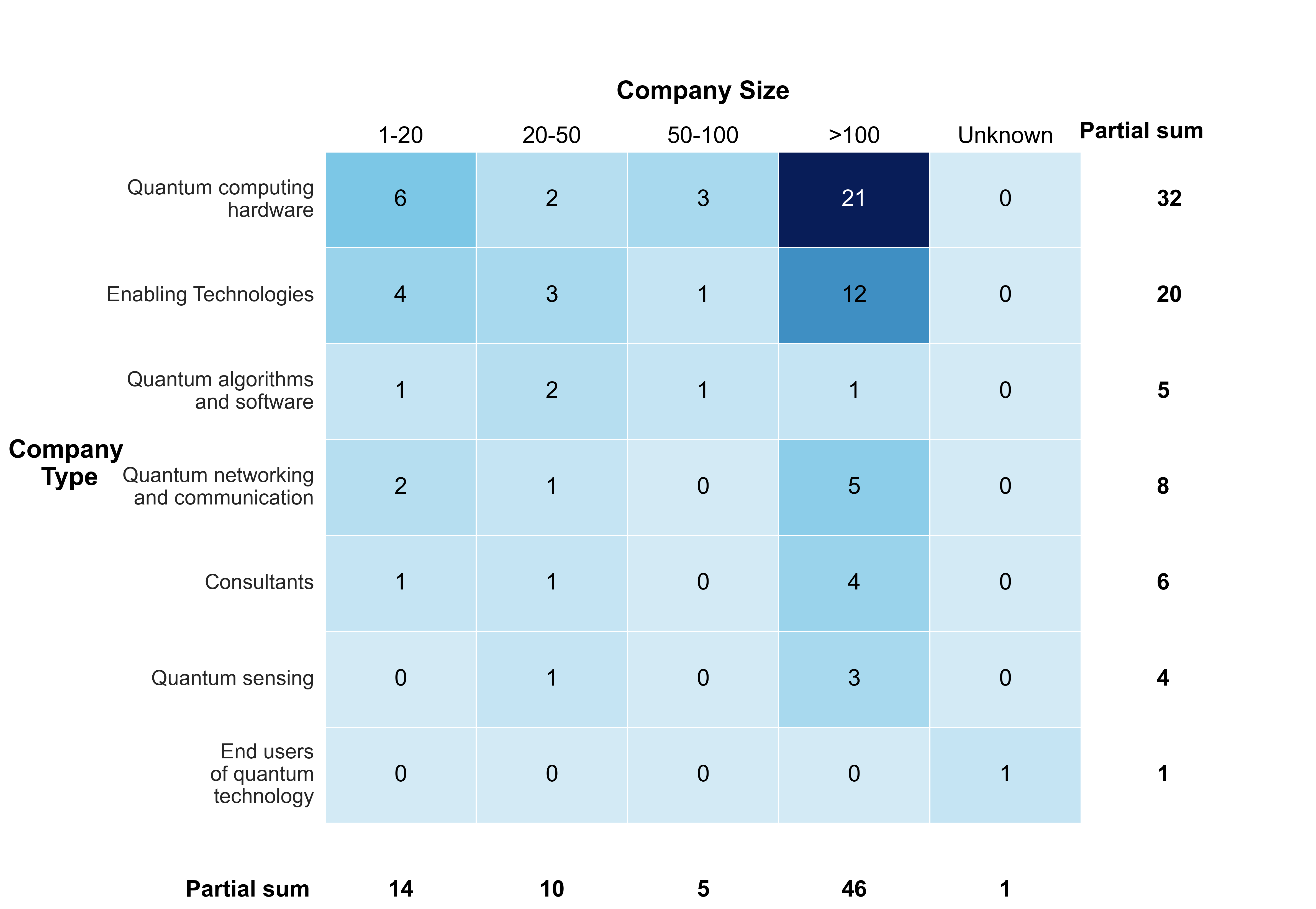}
  \caption{Distribution of interviewed companies by type and size.  We did 37 interviews with professionals at 22 distinct companies. We had a few interviewees from the same company and most companies self-reported participation in more than one type of activity. We adopted definitions consistent with prior literature \cite{fox2020preparing, el2025industry} for company types. Company size corresponds to the range of employees working in quantum-related technologies in these companies. \emph{Note: Figure 1 in the first version of this preprint included only a subset of the distribution of companies analyzed in this report. The figure has been updated to show the distribution type and size of all 37 interviews.}}
  \label{fig:categorization}
  
\end{figure}

Participants represented a range of perspectives and roles within the industry: 
\begin{itemize}
  \item Managers who provided a broad view of different roles in their company and associated knowledge, skills, and abilities (KSAs);
  \item Employees who described detailed tasks they engage in and associated KSAs.
\end{itemize}
Interviewees came from various company types and sizes as seen in Fig. \ref{fig:categorization}, which allowed us to have insights into workforce needs from a variety of job roles in the quantum industry.

\begin{questionbox}
\textbf{What experimental skills are needed for non-PhD roles in the quantum industry?}
\end{questionbox}

To answer our key question, we analyzed our interview data to identify the knowledge and skills needed for each individual position in our dataset. A total of 70 individual positions were identified and filtered to 22 requiring associate's, bachelor's or master's degrees. Initial analysis used skill categories from the AAPT Recommendations for the Undergraduate Physics Laboratory Curriculum framework \cite{2014aapt}, which we later refined to better reflect the pattern of skills found in our dataset.

\paragraph*{Note on terminology}\mbox{}\\
\textit{Job role} often refers to a job family or title that multiple people may hold across different companies, even if their responsibilities may vary somewhat. This report focuses on individual \textit{positions}, by which we mean a specific job at a particular company (e.g., quantum software developer at company X). When we use the term ``role'' in this report, it is for readability and should be interpreted as referring to a specific position unless noted otherwise.

\newpage
\tocsection{Results}

\subsection*{Experimental skills for quantum industry positions held by non-PhDs}
\addcontentsline{toc}{subsection}{Experimental skills for quantum industry positions held by non-PhDs}

\begin{takeawaybox}
Our study identifies a set of experimental skills listed in Tables \ref{tab:instrumentation_skills}, \ref{tab:computation_skills}, \ref{tab:experimental_design_skills}, \ref{tab:communication_skills} organized into four categories:
\begin{enumerate}
  \item \textbf{Instrumentation}: Operating, maintaining, and troubleshooting scientific and engineering instruments and systems.
  \item \textbf{Computation and data analysis}: Applying computational tools to acquire, process, and interpret data.
  \item \textbf{Experimental and project design}: Planning, executing, and managing experiments.
  \item \textbf{Communication and collaboration}: Sharing technical insights and coordinating with stakeholders.
\end{enumerate}
\end{takeawaybox}

The experimental skills identified from our data are listed by skill category in Tables \ref{tab:instrumentation_skills}, \ref{tab:computation_skills}, \ref{tab:experimental_design_skills}, \ref{tab:communication_skills}.

\begin{table}[H]
\centering
\caption{Instrumentation skills (I1–I9) for quantum industry positions held by non-PhDs.}
\label{tab:instrumentation_skills}

\renewcommand{\arraystretch}{1.1}
\setlength{\tabcolsep}{4pt}
\small

\begin{tabularx}{\textwidth}{l X}
\toprule
\textbf{} & \textbf{Instrumentation Skills} \\
\midrule
I1 & Able to operate basic test and measurement devices (e.g., power supplies, multimeters) \\
I2 & Having hands-on experience with electronic systems and components \\
I3 & Able to operate and fix laser systems \\
I4 & Having hands-on experience with optical systems and optical alignment \\
I5 & Able to operate interferometers to assess surface flatness \\
I6 & Able to work with ultra-low temperature apparatus \\
I7 & Able to work with vacuum systems \\
I8 & Able to do basic troubleshooting with an iterative and logical approach \\
I9 & General practical experimental skills (e.g. research experience, lab course) \\
\bottomrule
\end{tabularx}
\end{table}

\begin{table}[H]
\centering
\caption{Computation \& data analysis skills (D1–D6) for quantum industry positions held by non-PhDs.}
\label{tab:computation_skills}

\renewcommand{\arraystretch}{1.1}
\setlength{\tabcolsep}{4pt}
\small

\begin{tabularx}{\textwidth}{l X}
\toprule
\textbf{} & \textbf{Computation \& Data Analysis Skills} \\
\midrule
D1 & Able to interpret computational or measurement results to understand physical systems \\
D2 & Able to do some programming generally, without specifying whether it is for data processing, data representation, or other purposes \\
D3 & Able to select numerical algorithms or software tools based on problem requirements \\
D4 & Able to troubleshoot code \\
D5 & Able to use computers to collect and process data \\
D6 & Able to perform data analysis including uncertainty analysis \\
\bottomrule
\end{tabularx}
\end{table}

\begin{table}[H]
\centering
\caption{Experimental \& project design skills (E1–E10) for quantum industry positions held by non-PhDs.}
\label{tab:experimental_design_skills}

\renewcommand{\arraystretch}{1.1}
\setlength{\tabcolsep}{4pt}
\small

\begin{tabularx}{\textwidth}{l X}
\toprule
\textbf{} & \textbf{Experimental \& Project Design Skills} \\
\midrule
E1 & Able to integrate multiple scientific principles when designing experiments \\
E2 & Able to integrate multiple scientific principles to translate findings into practical applications \\
E3 & Able to design a procedure to test a model or hypothesis \\
E4 & Able to effectively plan and carry out experiments \\
E5 & Able to manage complex technical projects from formulation through testing to reporting \\
E6 & Able to perform basic systems engineering tasks, including understanding requirements and determining acceptable performance thresholds \\
E7 & Able to design and oversee construction of specialized scientific facilities, including cleanroom and chip fabrication facilities for quantum research \\
E8 & Able to critically read scientific literature to refine research questions or optimize an experimental design \\
E9 & Able to conduct an experiment in an open-ended and ill-defined context (e.g., working in ambiguity with emerging technology) \\
E10 & Able to persist through experimental setbacks, including adapting approaches until reliable outcomes are achieved \\
\bottomrule
\end{tabularx}
\end{table}

\begin{table}[H]
\centering
\caption{Communication \& collaboration skills (C1–C5) for quantum industry positions held by non-PhDs.}
\label{tab:communication_skills}

\renewcommand{\arraystretch}{1.1}
\setlength{\tabcolsep}{4pt}
\small

\begin{tabularx}{\textwidth}{l X}
\toprule
\textbf{} & \textbf{Communication \& Collaboration Skills} \\
\midrule
C1 & Able to prepare and deliver presentations to multiple stakeholders \\
C2 & Able to collaborate effectively with others \\
C3 & Able to create and maintain clear and up-to-date technical documentation for experimental setups and procedures \\
C4 & Able to write reports and proposals \\
C5 & Able to critically assess both their own work and that of others to identify strengths and limitations \\
\bottomrule
\end{tabularx}
\end{table}

\subsection*{Overview of non-PhD quantum industry positions}
\addcontentsline{toc}{subsection}{Overview of non-PhD quantum industry positions}

\vspace{0.2cm}

\begin{takeawaybox}
\begin{itemize}
    \item The \textbf{22 non-PhD positions} identified in our dataset span a wide range of job titles and disciplinary backgrounds, from highly hands-on technical positions (e.g., \textit{lab technician, fabrication engineer, assembly technician}) to engineering roles (e.g., \textit{research engineer, quantum engineer, nanofabrication engineer}) to more business-focused roles (e.g., \textit{quantum business development specialist, product manager}). 
    \vspace{0.2cm}
    \item The unifying feature of most non-PhD positions is their \textbf{reliance on experimental skills}: whether the day-to-day tasks involve assembly of optical systems, troubleshooting, developing research software, or presenting results, most rely heavily working knowledge and skills in \textbf{instrumentation}, \textbf{computation and data analysis}, \textbf{experimental and project design}, and \textbf{communication and collaboration}. 
\end{itemize}
\end{takeawaybox}

\begin{table}[H]
\centering
\caption{Overview of the 22 individual positions, including education, discipline, quantum expertise, and mapping to the four experimental skill categories (refer to Tables \ref{tab:instrumentation_skills}, \ref{tab:computation_skills}, \ref{tab:experimental_design_skills}, \ref{tab:communication_skills} for exact skill).}
\label{tab:nonphd_jobs}

\scriptsize
\renewcommand{\arraystretch}{1.35}
\setlength{\tabcolsep}{3pt}

\begin{adjustbox}{max width=\textwidth, center}
\rowcolors{2}{tabstripe}{white}
\begin{tabular}{|C{0.16\textwidth}|
                C{0.12\textwidth}|
                C{0.18\textwidth}|
                C{0.14\textwidth}|
                S{0.16\textwidth}|
                S{0.16\textwidth}|
                S{0.16\textwidth}|
                S{0.16\textwidth}|}
\hline
\rowcolor{tabheadbg}
\textbf{\color{brandnavy} Title of Individual Positions} &
\textbf{\color{brandnavy} Education} &
\textbf{\color{brandnavy} Discipline} &
\textbf{\color{brandnavy} Quantum Expertise} &
\textbf{\color{brandnavy} Instrumentation} &
\textbf{\color{brandnavy} Computation \& Data Analysis} &
\textbf{\color{brandnavy} Experimental \& Project Design} &
\textbf{\color{brandnavy} Communication \& Collaboration} \\
\hline

Lab technician & Associate, Bachelor & Engineering (Eng.), Other & Aware & I9 & D2, D5 &  &  \\ \hline
Fabrication engineer & Bachelor & Physics, Math & Conversant, Proficient & I1–I8 & D1, D5 & E1, E2, E5, E10 & C5 \\ \hline
System operator & Bachelor & Physics, Engineering & Conversant & I1, I8 & D1 & E1, E2, E6 & C1, C3 \\ \hline
Photonics assembly technician & Bachelor & Physics, Engineering & Conversant, Proficient & I2, I4, I9 & D2, D6 & E5 & C5 \\ \hline
Assembly technician & Bachelor & Physics & Aware, Conversant & I2, I3, I4, I9 & D2, D6 & E5 & C1, C2, C5 \\ \hline
Quantum R\&D engineer & Bachelor & Engineering Physics & Proficient & I2, I4, I8 & D2 & E3, E4, E6, E8 & C1, C2, C3 \\ \hline
Laser and optics engineer & Bachelor & Physics, Chemistry & Proficient & I4, I5, I8, I9 & D2 &  &  \\ \hline
Quantum software developer 1 & Bachelor & Computer Engineering & Conversant &  & D1, D2, D3, D4 & E10 & C1, C2, C3, C5 \\ \hline
Quantum software developer 2 & Bachelor & Computer Science (CS) & Conversant &  & D1, D4 &  & C1, C2, C5 \\ \hline
Quantum software engineer 1 & Bachelor & CS, Engineering, Math & None, aware &  & D1, D2, D4 & E10 & C1, C2, C5 \\ \hline
Quantum software engineer 2 & Bachelor & CS, Physics, Math & Conversant &  & D1, D2, D4 &  & C1, C2, C5 \\ \hline
Chief Operating Officer & Bachelor & Engineering, Other & Aware & I8 &  & E5 & C1 \\ \hline
Construction specialist & Bachelor & Other & Aware &  &  & E5, E7 & C1 \\ \hline
Research engineer & Bachelor, Master & Physics, Chemistry, CS, Eng. & Conversant & I3, I4, I8 &  & E5 & C2 \\ \hline
Research scientist / quantum systems engineer & Bachelor, Master & Physics, Thermal Eng., Mech. Eng. & Proficient & I8, I9 &  & E10 & C2, C5 \\ \hline
Algorithm developer 1 & Bachelor, Master & CS, Physics, Math, Other & Proficient &  & D1, D2, D3, D4, D6 & E3, E8, E10 & C1, C2, C5 \\ \hline
Algorithm developer 2 & Bachelor, Master & Physics, CS & Proficient &  & D1, D3, D4 &  & C5 \\ \hline
Quantum engineer & Bachelor, Master & Engineering, Eng. Physics & Conversant, proficient & I3 &  & E6 &  \\ \hline
Test and integration engineer & Master & Electrical Engineering & Conversant & I5, I6, I7, I8 & D3 & E6, E9 & C1, C2, C5 \\ \hline
Nanofabrication engineer & Master & Physics & Aware & I8 &  &  & C5 \\ \hline
Quantum business development specialist & Master & Physics & Expert &  & D6 & E5 & C1, C4 \\ \hline
Product manager & Master & Engineering & Conversant &  &  & E5 & C1, C5 \\ \hline

\end{tabular}
\end{adjustbox}

\vspace{3pt}
\begin{minipage}{0.98\textwidth}\footnotesize
\textit{Legend:} ``Other'' denotes that no specific discipline is required according to the interviewee. ``Quantum Expertise'' describes the level of quantum knowledge the interviewee believes the position requires. Definitions provided to interviewees were adapted from the CUbit Quantum Initiative's report on the quantum workforce \cite{CUbit2024Roadmap}:\\
\emph{Quantum Aware}: those who are new to quantum concepts and simply have general awareness of the subject matter; \\
\emph{Quantum Conversant}: those may or may not hold a traditional four-year degree, but should be able to have a conversation in quantum information science topics; \\
\emph{Quantum Proficient}: those who may hold physics or engineering degrees in quantum-specific topics at either the undergraduate or master’s degree levels; \\
\emph{Quantum Expert}: those who may typically hold a PhD in physics, engineering, computer science, or other subjects connected to the basic study of quantum phenomena.\\
\end{minipage}
\end{table}

\newpage
\subsection*{Patterns of experimental skills among non-PhD professionals in the quantum industry}
\addcontentsline{toc}{subsection}{Patterns of experimental skills among non-PhD professionals in the quantum industry}

\begin{figure}[h!]
  \centering
  \includegraphics[width=0.9\linewidth]{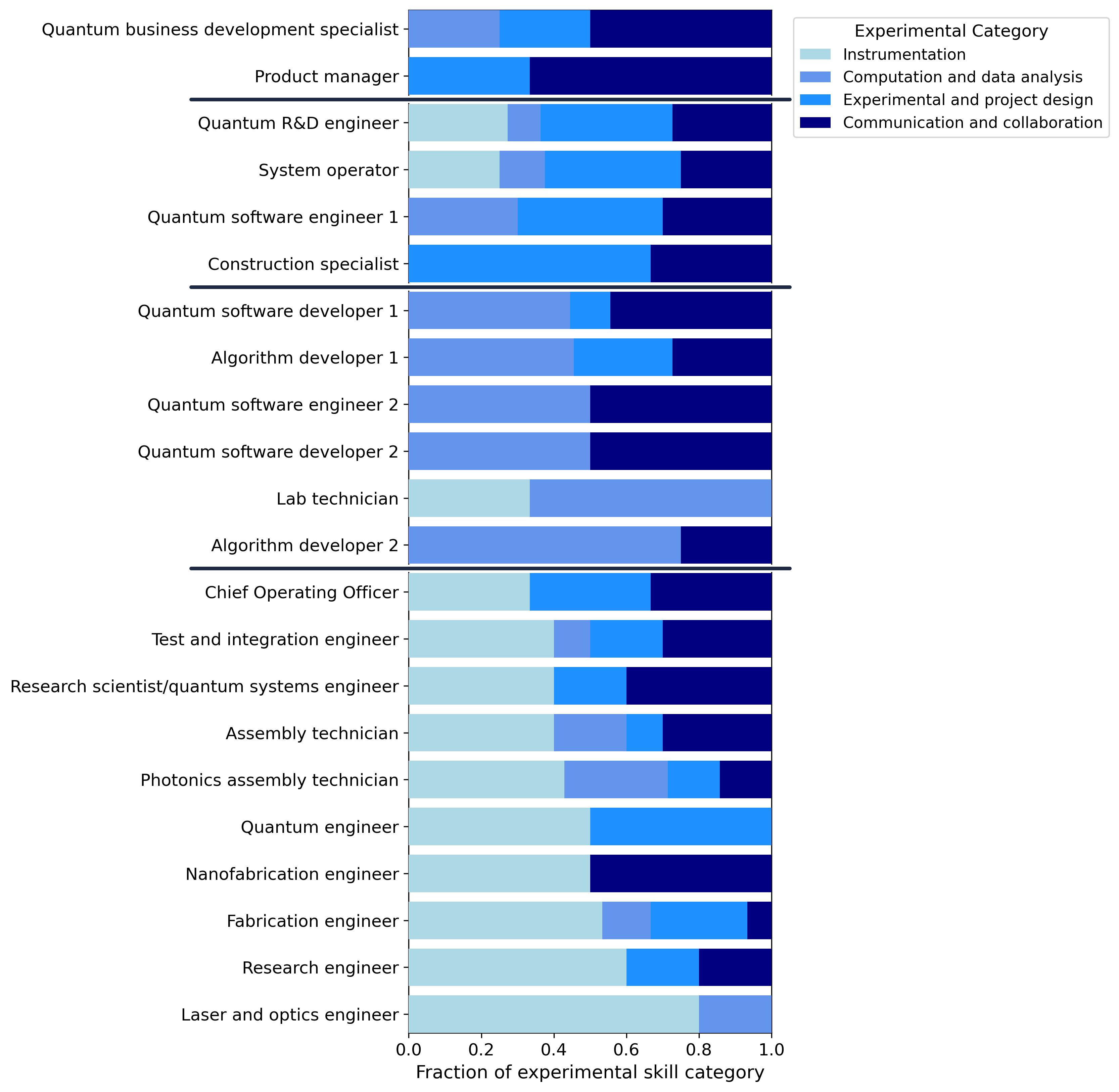}
  \caption{Distribution of experimental skills for each position grouped by dominant experimental skills category.
  The individual positions serving as starting and endpoint points for each category are:
  Chief Operating Officer $\rightarrow$ laser \& optics engineer  emphasize \textbf{instrumentation} skills.
  Algorithm developer~2 $\rightarrow$ software developer~1 emphasize \textbf{computation and data analysis} skills.
  Construction specialist $\rightarrow$ quantum R\&D engineer emphasize \textbf{experimental and project design} skills.
  Product manager $\rightarrow$ quantum business development specialist emphasize \textbf{communication and collaboration} skills.}
  \label{fig:experimentalskills}
\end{figure}

\begin{takeawaybox}
\begin{itemize}
    \item The \textbf{22 positions} require a mix of the four skill categories as illustrated in Fig. \ref{fig:experimentalskills}, though certain positions place greater emphasis on particular skills. Hardware focused positions emphasize \textbf{instrumentation} skills. Software and algorithm positions emphasize \textbf{computation and data analysis} skills. Customer and product development positions emphasize \textbf{experimental and project design}, and \textbf{communication and collaboration} skills.
    \item These patterns reflect both the technical nature of the work in these positions and the \textbf{typical educational levels} of individuals filling them. Bachelor's level positions tend to focus on building and operating experimental setups, while master's level positions focus more on planning and optimizing those setups, as well as communicating effectively with various stakeholders.
\end{itemize}
\end{takeawaybox}

\subsubsection*{Instrumentation}
Instrumentation skills emerged as  dominant desirable skills across a wide range of roles in our data and serve as a key differentiator between positions open to different degree levels. These skills range from foundational skills, such as operating basic test and measurement devices, to more specialized expertise, such as working with vacuum systems. 

Some positions with an instrumentation focus are accessible to both bachelor's or master's degree holders such as the \textit{quantum engineer} position. However, positions that require only a bachelor's degree tend to emphasize familiarity with common laboratory tools and systems. For instance, when discussing the knowledge and skills needed for a \textit{fabrication engineer}, our interviewee explains the importance of being able to operate basic test and measurement devices. They say:
\begin{quote}
Understanding how a multimeter works and what impedance is — these basic things are helpful.
\end{quote}
Other essentials skills at this educational level include operating interferometers to assess surface flatness and be able to do basic troubleshooting of equipment.
At the master's level, instrumentation skills shift towards integrating subsystems, and using advanced optomechanical and design tools. For example, when discussing the knowledge and skills needed for the \textit{test and integration engineer} position, our interviewee highlights the need for candidates with more hands-on experience to build and maintain the structures that support optical systems. They explain:

\begin{quote}
The big one that we struggled to find are people with the optomechanical knowledge and tools such as CAD, [Computer-Aided Design] SolidWorks or Onshape that support making a system.
\end{quote}

\medskip
The breadth of these roles illustrates how proficiency with scientific instruments and systems opens multiple opportunities. 
For \textit{bachelor’s graduates}, hands-on experience with laboratory equipment provides a foundation for technical positions focused on assembly, maintenance, and troubleshooting. 
For \textit{master’s graduates}, more advanced and specialized instrumentation skills enable individuals to take on system integration and optimization responsibilities.

\vspace{0.6cm}

\subsubsection*{Computation and data analysis}
Computation and data analysis are critical for roles focused on software development. In our dataset, these positions are largely accessible to individuals with a \textit{bachelor’s degree} and emphasize the ability to interpret computational or measurement results to understand physical systems. 

In these roles, employees are expected to have general programming skills that enable them to write code. When discussing knowledge and skills needed for the \textit{quantum software developer 1} position, our interviewee states that familiarity with common programming environments is a baseline expectation:

\begin{quote}
It's expected that people are familiar with either Python or Conda, the Jupyter Notebook environment that type of R\&D set.
\end{quote}

Beyond basic programming, these positions require the ability to select appropriate numerical algorithms or software tools based on problem requirement. Additionally, in these roles, employees need to perform data analysis, including uncertainty analysis, to inform experimental results or develop software tools that support quantum research and engineering. By combining programming expertise and data interpretation, individuals in these roles provide the software infrastructure and analytical skills that bridge the gap between experimental operations and computational modeling.

\vspace{0.6cm}

\subsubsection*{Experimental and project design}
Several positions place experimental and project design skills at the core of their responsibilities, which require individuals to plan, execute, and manage experiments or experimental workflows. These positions emphasize the ability to design procedures to test models or hypotheses, and effectively plan and carry out experiments.

Positions emphasizing these skills are filled by bachelor's level graduates who must be able to integrate multiple scientific principles to translate findings into practical applications. These individuals balance theoretical concepts with practical considerations.  For example, when discussing knowledge and skills needed for the \textit{quantum R\&D engineer} position, our interviewee describes that part of managing a complex technical projects includes:

\begin{quote}
 Translating a quantum experimental plan into requirements for a control system such as resolution, or number of channels of a certain output type. 
\end{quote}

\medskip
 Employees in these roles that emphasize experimental and project design also highlight the importance of adaptability and problem-solving when conducting experiments in open-ended and ill-defined contexts. Individuals in these positions play a key role in shaping experimental processes and adapting them to evolving project needs.

\vspace{0.6cm}

\subsubsection*{Communication and Collaboration}
Positions that emphasize communication and collaboration are essential for bridging technical work with strategic decision-making. 
These positions tend to be filled by individuals with a \textit{master’s degree}, which reflects the advanced experimental skills needed to communicate and collaborative effectively with multiple stakeholders. For example, when discussing the skills needed for the \textit{product manager} position, our interviewee highlights how a deep understanding of experiments helps in navigating conflicting technical requirements:

\begin{quote}
 When I have a customer interaction, when the customer stipulated two seemingly unrelated technical requirements, but in actual implementation, when you understand how the experiment works, they are actually contradictory to each other, you have to bring it up to the customer and say, "Hey, if we're going to implement to fulfill your requirement A, we need to do this. When doing this, that's actually contradicting requirement B that you just stipulated, even though on the surface, these two don't seem related to each other." That's definitely a very important skill set that I would say that really helps out in my line of work.
\end{quote}
In addition to external communication, these positions require strong internal communication practices, such as creating and maintaining clear and up-to-date technical documentation for experimental setups and procedures.  Professionals in these positions focus on translating complex technical results into actionable business strategies that ensure projects move forward smoothly. Ultimately these roles that emphasize communication and collaboration skills serve as a vital link between experimental teams and strategic leadership.

\newpage
\tocsection{Recommendations and conclusions}

\begin{takeawaybox}

\begin{itemize}
  \item Non-PhD roles in the quantum industry are unified by an emphasis on experimental skills across four areas: \textbf{instrumentation}, \textbf{computation and data analysis}, \textbf{experimental and project design}, and \textbf{communication and collaboration}.
  \item Bachelor’s roles emphasize a range of skills in \textbf{instrumentation}, \textbf{computation and data analysis}, and \textbf{experimental and project design}, while master’s roles place greater weight on \textbf{communication and collaboration} on top of the other three skill categories.
\end{itemize}
\end{takeawaybox}

These findings point to concrete steps educators can take to align QISE courses and program with workforce needs:

\begin{enumerate}

   \item \textbf{Prioritize the discussion of hardware in QISE theory courses}: Undergraduate programs should emphasize the theoretical foundations of quantum hardware within QISE courses  to better prepare students for entry-level positions in the quantum industry. 
   Currently, many introductory QISE courses focus solely on abstract theoretical concepts with limited time spent on theory underlying the experiments and devices \cite{pina2025landscape}. 
   Thus, shifting the focus of these courses to include discussions on theory behind hardware components will better align QISE education with industry needs. 
   \item \textbf{Increase experimental training through instructional labs}: Many of the experimental skills needed for non-PhD positions in the quantum industry overlap with general experimental skills that are valuable across a range of career paths. Many of these  skills are already part of the learning goals of existing lab courses, which makes this recommendation achievable with minimal additional resources. Increasing opportunities to engage in experimental skill development within existing labs would better prepare students for positions in the quantum industry, while simultaneously strengthening preparation for other career paths.
   \item \textbf{Include professional skills in undergraduate QISE education}: Employers consistently identify communication and collaboration as critical skills. These skills should be embedded within QISE courses and programs to ensure students are prepared for industry roles. \end{enumerate}

\subsection*{Limitations}

Our findings should be interpreted with the following limitations in mind:

\begin{itemize}
  \item \textit{Sample and scope:} We interviewed 37 professionals at 22 companies. Thus, the sample is not intended to be representative of all roles in the quantum industry.
  \item \textit{Time-bounded snapshot:} The report presents the first set of findings based on interviews conducted with quantum industry professionals in the first half of 2025. Workforce needs may evolve as quantum technologies becomes more mature.
\end{itemize}

\subsection*{Future work}

Building on these initial results and addressing the limitations above, we plan to:
\begin{itemize}
   \item Further refine the experimental skills taxonomy as we collect more data;
  \item Provide a detailed account of the extent and type of QISE content knowledge expected at the bachelor’s and master’s levels as we conduct more analysis.
\end{itemize}
\newpage
\tocsection{Acknowledgments}
Thank you to the quantum industry managers and employees who participated in our interviews. 

This material is based on work supported by the National
Science Foundation under Grant Nos. PHY-2333073 and PHY-2333074.

This material is also based on work supported by the Army Research Office and was accomplished under Award Number: W911NF-24-1-0132. The views and conclusions contained in this document are those of the authors and should not be interpreted as representing the official policies, either expressed or implied, of the Army Research Office or the U.S. Government. The U.S. Government is authorized to reproduce and distribute reprints for Government purposes notwithstanding any copyright notation herein.

\vspace{0.9\baselineskip}
{\sffamily\bfseries\color{brandnavy} About the project}\\
This report is part of a collaborative project between researchers at the University of Colorado Boulder \& Rochester Institute of Technology to advance quantum information science education and strengthen the quantum workforce.  
Learn more about the broader effort at
\href{https://www.rit.edu/quantumeducationandworkforce/}{rit.edu/quantumeducationandworkforce}.

\vspace{0.9\baselineskip}
{\sffamily\bfseries\color{brandnavy} Contact information}\\
For questions about this report or the broader project, please contact the principal investigators:

\noindent\hfill
\begin{minipage}{0.9\linewidth}
\centering
\begin{tabular}{@{}p{0.45\linewidth}p{0.45\linewidth}@{}}
Heather J.~Lewandowski & Benjamin M.~Zwickl \\
University of Colorado Boulder & Rochester Institute of Technology \\
\href{mailto:lewandoh@colorado.edu}{lewandoh@colorado.edu} &
\href{mailto:bmzsps@rit.edu}{bmzsps@rit.edu} \\
\end{tabular}
\end{minipage}
\hfill\mbox{}

\vspace{0.9\baselineskip}
{\sffamily\bfseries\color{brandnavy} Suggested citation}\\
Shams El-Adawy, A.\,R. Pi\~na, Benjamin M.\,Zwickl, \& H. J. Lewandowski (October 2025).
\emph{Experimental Skills for Non-PhD Roles in the Quantum Industry}
(Quantum Workforce Report Series, Report 1). University of Colorado Boulder \& Rochester Institute of Technology.

\newpage

\printbibliography
\end{document}